\documentclass[debug,overfull]{epl2}

\usepackage{graphicx}
\usepackage{epsfig}
\usepackage{amssymb}
\usepackage{amsmath}
\usepackage{amsfonts}
\usepackage{graphicx}

\title{Analogue Casimir Radiation using an Optical Parametric Oscillator}
\author{F.X. Dezael\inst{1} and A. Lambrecht\inst{1}} \institute{\inst{1} Laboratoire Kastler Brossel,
CNRS, ENS, UPMC - Campus Jussieu case 74, 75252 Paris, France}


\abstract{ We establish an explicit analogy between the dynamical
Casimir effect and the photon emission of a thin non-linear crystal
pumped inside a cavity. This allows us to propose a system based on
a type-I optical parametric oscillator (OPO) to simulate a cavity
oscillating in vacuum at optical frequencies. The resulting photon
flux is expected to be more easily detectable than with a mechanical
excitation of the mirrors. We conclude by comparing different
theoretical predictions and suggest that our experimental proposal
could help discriminate between them.}

\begin{document}
\maketitle

Any mirror placed in quantum vacuum experiences fluctuations of the
vacuum radiation pressure. When moving with non-uniform
acceleration, these fluctuations give rise to a dissipative force,
opposing itself to the mirror's motion \cite{fulling}. As a
counterpart, the mirror should emit photons into the free field
vacuum because of energy conservation. Although predicted 30 years
ago in the seminal paper by Fulling and Davies, this so-called
dynamical Casimir effect has not yet been observed experimentally, mainly
because the order of magnitude of the predicted photon
flux is very small.

Different attempts have been made in the past to render
the dynamical Casimir effect observable, for instance by exploiting
the resonant enhancement of radiation inside an oscillating cavity
\cite{Lambrecht1996,Lambrecht1998,Crocce2002,Uhlmann2004} or by
amplifying the Casimir signal with a sample of superradiant
ultracold alkali-metal atoms\cite{Onofrio06}. Another possibility
consists in effectively simulating the displacement of a mirror, for
example by rapidly changing or modulating the skin depth of a
semiconductor. This has first been discussed in \cite{Yablonovich89}
and \cite{Lozovik95} for linear and non-linear acceleration
respectively and later-on been implemented in an experimental
proposal \cite{Padova}. Another more recent paper has proposed the
generation of Casimir radiation via the coupling of a qubit to a
cavity \cite{Carusotto09b}. Noteworthy related effects based on the
same physical phenomenon are the fibre-optical analogue of the event
horizon where a light pulse mimics a moving medium \cite{Philbin08}
or the emission of acoustic Hawking radiation via phonon modulation
in Bose-Einstein condensates \cite{Carusotto09a}.

In this Letter, we propose to generate the analogue of Casimir
radiation inside a type-I optical parametric oscillator (OPO). Such
a device is commonly used in Quantum Optics\cite{Yariv1968}, but
here we intend to drive it in a specific Casimir-like emission
regime. We first establish an analogy between photon emission of a
coherently pumped non-linear crystal of type $\chi^{(2)}$ inside a
cavity and the photon creation via the dynamical Casimir effect. By
giving the equivalence relations between the parameters of
mechanical motion and the characteristics of the crystal and pump
field, we show that the non-linear interaction results in an
\emph{apparent} motion of the mirrors for the electromagnetic field.
We give an analytical expression for the photon flux emitted by the
model system and discuss orders of magnitude of this Casimir-like
radiation, which we find to be easily achievable in a standard
experiment. This is due to the fact that our model system avoids
mechanical motion, which is the limiting factor in the oscillating
cavity proposals, but uses instead an apparent motion of the mirrors
for the field. It rejoins insofar experimental proposals where the
mechanical motion is replaced by the optical modulation of the
mirrors skin depth \cite{Yablonovich89,Lozovik95,Padova}.

Based on the above mentioned analogy, we then establish a link
between the recently introduced concept of time refraction
\cite{mendonca2005,mendonca2008} and the time varying refractive
properties of the pumped crystal, leading to an alternative
expression for the photon flux emitted by a cavity oscillating in
vacuum. Such as many others this expression results in an
exponential growth of emitted photons at all times. This is not only
the case when perfectly reflecting mirrors are considered
\cite{Dodonov1995,Uhlmann2004}, but also when a damping coefficient
$\gamma$ for the energy is introduced
\cite{Schaller2002,mendonca2008}. We compare these results with our
predictions, obtained within the scattering approach where the
fields' transformations are directly evaluated on mirrors with
finite reflection coefficients
\cite{Lambrecht1996,Lambrecht1998,EPL1998}. The latter procedure
leads automatically to a stationary regime with a finite number of
photons emitted inside the cavity and constant flux emitted outside.
We show that the models predicting an exponential growth are only
valid in the short time limit and lead back to our results when
accounting for a detailed balance between photons emitted by the
mirrors and photons leaving the cavity due to their finite
reflectivity.

The model system that we will consider in analogy with dynamical
Casimir radiation is a type-I optical parametric oscillator (OPO)
with a $\chi^{(2)}$ non-linear crystal of length $l$ such as
schematically shown in Fig. 1.
\begin{figure}[tbh]
\begin{center}
\includegraphics[width=12cm]{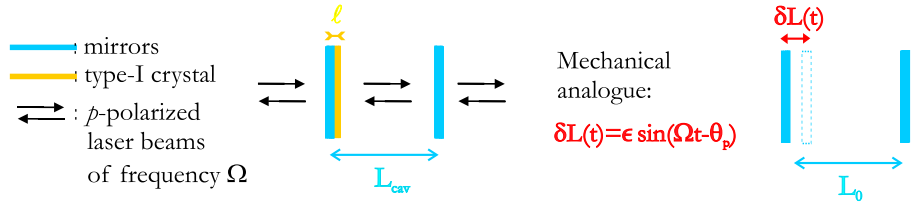}
\caption{Type-I Optical Parametric Oscillator with one thin crystal
slab stuck on the left mirror of the cavity, and equivalent
oscillating cavity.}
\label{fig1}
\end{center}
\end{figure}
As usually this system is conveniently described using Maxwell
equations and the coupling of the field with the atoms inside the
dielectric crystal. If the pumping field is intense, a non-linear
polarization vector $\vec{P}^{(NL)}$ must be added to the usual
linear contribution
\begin{eqnarray}
P_{i}[\vec{r},\omega]&=&\varepsilon_{0}
\chi^{(1)}_{i,j}[\omega] E_{j}[\vec{r},\omega]+P^{(NL)}_{i}[\vec{r},\omega]
\label{vect pol total}\\
P^{(NL)}_{i}[\vec{r},\omega] &=& \frac{\varepsilon_{0}}{2}  \int
\frac{d\omega'}{2\pi} \chi^{(2)}_{i,j,k}\big[\omega;
\omega-\omega',\omega'\big]
E_{j}[\vec{r},\omega-\omega']E_{k}[\vec{r},\omega'],
\label{expression de la polarisation non-linéaire chi(2) 3D}
 \end{eqnarray}
where we have used the Fourier representation of fields. $i$ and $j$
stand for the projection of vectors on the 3 orthogonal spatial
axes $x,y,z$, and we use the summing convention on
repeated indices. $\chi^{(1,2)}$ are respectively the
linear and non-linear susceptibility tensor. $\chi^{(1)}$ is
diagonal if we choose $x,y,z$ parallel to the proper directions of
the crystal,
$\chi^{(1)}_{i,j}[\omega]=(n_{i}(\omega)^{2}-1)\delta_{i,j}$,
$n_{i}$ being the refraction index of the crystal in direction $i$.
For non centro-symmetrical crystals and reasonable pumping fluxes,
the non-linear polarization vector exhibits dominantly a
$\chi^{(2)}$ effect. For simplicity, we will consider that the
crystal is pumped with a single laser beam of frequency $\Omega$,
propagating rightwards along the $x$-axis and linearly polarized
along the $p$-axis ($p=y,z$). If we denote $\Phi$ its photon flux
and $\theta_{p}$ its phase at $x=0$, the pumping field inside the
crystal writes
$ E_{p}(\vec{r},t)= E_0 \, e^{\iota\theta_{p}} e^{-\iota
\Omega\big(t-n_{_{p}}(\Omega)x/c\big)} + c.c.$ where
$E_0=\sqrt{\frac{\hbar\Omega\Phi}{2\varepsilon_{0}cAn_{p}(\Omega)}}$
is the pump field's amplitude, and $\iota$ is the imaginary unit. The constant $A$ represents the
transverse extension of the beams.

We consider a type I interaction inside the crystal. In this case,
if we denote '$s$' the transverse polarization perpendicular to
'$p$', only the tensorial elements
$\chi^{(2)}_{s,s,p}=\chi^{(2)}_{s,p,s}$ of equation (\ref{expression
de la polarisation non-linéaire chi(2) 3D}) contribute to the
non-linear polarization. By assuming that the crystal is
non-absorbant and dispersion-free in the spectral range of interest,
we can replace the crystal's linear susceptibility and second order
non-linear susceptibility by their average values
$\overline{\chi^{(1)}}$ and $\overline{\chi_{s,s,p}^{(2)}}$. In this
case, the '$s$' component of the polarization vector writes
$P_{s}(\vec{r},t) = \; \varepsilon_{0} \, \tilde{\chi}^{(1)}(x,t) \,
E_{s}(\vec{r},t)$, with
$\tilde{\chi}^{(1)}(x,t)=\overline{\chi^{(1)}} + \,
\kappa\,\sin\big[\Omega (t - \overline{n}_{p}x/c) -
\theta_{p}\big]$, $\overline{n}_{p}$ being the mean
refraction index of the crystal in direction '$p$' and $\kappa=
\sqrt{\frac{\hbar \Omega \Phi}{2\varepsilon_{0}c A
\overline{n}_{p}}} \; \overline{\chi_{s,s,p}^{(2)}}$. The crystal
thus behaves as a linear dielectric medium in direction '$s$', with
an effective refraction index $\tilde{n}_{s}(x,t)$ depending on
space-time coordinates $x$ and $t$: $\tilde{n}_{s}(x,t)^{2} =
\tilde{\chi}^{(1)}(x,t) + 1$. The typical order of magnitude of
$\kappa$ evaluates to $\kappa\simeq 10^{-5}$, if we consider pump
beams of power $\hbar \Omega \Phi\simeq 1$W, focalized over an area
$A\simeq 10^{-10}$m$^{2}$ inside a crystal of mean refraction index
$n\simeq1$ and non-linear susceptibility $\chi^{(2)}\simeq
10^{-11}$m.V$^{-1}$. We can thus safely expand the effective
refraction index to first order in $\kappa$:
\begin{eqnarray}
\tilde{n}_{s}(x,t) \simeq \overline{n}_{s} +
\frac{\kappa}{2\overline{n}_{s}} \,\sin\big[\Omega (t -
\overline{n}_{p}x/c) - \theta_{p}\big] \label{effective refr index
first order}
\end{eqnarray}

If the unperturbed length $l$ of the crystal is small compared to
the pump's wavelength $\Lambda=2\pi c/\Omega$ (say $l\simeq 0.1
\Lambda$), the spatial dependance of
$\tilde{n}_{s}(x,t)$ is not significant, and we can define an
average time-dependent refraction index $n_{s}(t) =\overline{n}_{s} + \frac{\epsilon_\textrm{opt}}{l}\sin\big[\Omega
t-\theta\big]$ with
\begin{equation}
\epsilon_\textrm{opt} \simeq \frac{l}{2\,\overline{n}_{s}}
\,\sqrt{\frac{\hbar \Omega \Phi}{2\varepsilon_{0}c A
\overline{n}_{p}}} \; \overline{\chi_{s,s,p}^{(2)}},
\label{epsilon
approx}
\end{equation}
while $\theta=\theta_{p}+\Omega
\overline{n}_{p}\{x_{0}+l/2\}/c -\pi/2$ if the crystal is located
between positions $x_{0}$ and $x_{0}+l$. The result of pumping can
then be seen as a periodic modulation of the effective length over
which the $s$-polarized fields propagate inside the crystal, that is
$l_{s}(t)\simeq n_{s}(t)\times l = \overline{n}_{s} l +
\epsilon_\textrm{opt} \sin\big[\Omega t-\theta\big] $. The frequency
of this modulation corresponds to the pumping frequency $\Omega$,
and its amplitude is given by (\ref{epsilon approx}).

Suppose now that we use a $p$-polarized laser beam of frequency
$\Omega$ to pump the system sketched in figure \ref{fig1}. Then the
temporal change in the refractive index results in a modulation of
the cavity length for $s$-polarized fields which is equivalent to an
apparent motion of the mirrors: $L(t)=L_{0}+\delta L(t)$, with
$L_{0}=L_{cav}+l\,\big(\overline{n}_{s}-1\big)$ and $\delta L(t)=l
\times \delta n_{s}(t)$. This modulation writes
\begin{equation}
L(t) = L_{0} + \epsilon_\textrm{opt} \sin\big[\Omega t-\theta\big],
\label{L(t) cavité pompage <-}
\end{equation}
meaning that optical parametric oscillators (OPO) can indeed
reproduce changes in boundary conditions equivalent to those
generating the dynamical Casimir effect. Accordingly we expect the
OPO model system to amplify the parametric fluorescence of the
crystal in a Casimir-like oscillation regime entailing the creation
of pairs of $s$-polarized correlated photons from vacuum, with
frequencies $\omega$ and $\omega '$ satisfying energy conservation:
$\hbar(\omega+\omega ')=\hbar \Omega$. The resulting signals are
resonantly enhanced when $\omega$ and $\omega '$ correspond to
cavity modes.

As we will discuss more precisely in a forthcoming paper, a single field mode of frequency $\omega$, propagating right- or leftward ($\overrightarrow{A}_\textrm{in}$ and $\overleftarrow{A}_\textrm{in}$) undergoes the following analogous transformation when interacting either with the composed "mirror-crystal" system, pumped at frequency $\Omega$, or with a mirror mechanically oscillating at frequency $\Omega$ around a mean position $x_{0}$:
\begin{eqnarray}
&&\overrightarrow{A}_\textrm{out}(\omega)=\sqrt{1-r} \, \overrightarrow{A}_\textrm{in}(\omega)+ \sqrt{r} \, e^{-\iota x_0 \omega/c} \overleftarrow{A}_\textrm{in}(\omega)  \\
&&+ \sqrt{r} \, e^{-\iota x_0 \omega/c} \bigg [ e^{\iota
\theta}\big(1-\frac{\omega}{\Omega}\big) \beta_\textrm{opt/mech}
\overleftarrow{A}_\textrm{in}(\omega-\Omega) +e^{-\iota
\theta}\big(1+\frac{\omega}{\Omega}\big) \beta_\textrm{opt/mech}
\overleftarrow{A}_\textrm{in}(\omega+\Omega) \bigg].\nonumber
\label{fieldtrafo}
\end{eqnarray}
$r=e^{-2 \rho}$ is the mirror's reflection coefficient. For
simplicity, it is chosen to be frequency independent and equal for
both mirrors of the cavity. The only difference between
mechanical and non-linear optical excitation lies in the expression
of the parameter $\beta_\textrm{opt/mech}$: for the mechanical
excitation it corresponds to the mirror's maximum velocity
$v_\textrm{max}$ with respect to the speed of light $c$ while for
the non-linear optical interaction it is a function of
the crystal's non-linearity and of the pump field's intensity and
frequency
\begin{eqnarray}
\beta_\textrm{mech} =
\frac{v_\textrm{max}}{c}=\frac{\epsilon_\textrm{mech}\Omega
}{c}\quad\quad \beta_\textrm{opt} = \frac{l \Omega}{2 c
\overline{n}_{s}}\sqrt{\frac{\hbar \Omega\Phi}{2\varepsilon_{0}c A
\overline{n}_{p}}} \overline{\chi_{s,s,p}^{(2)}} =
\frac{\epsilon_\textrm{opt} \Omega}{c}. \label{beta}
\end{eqnarray}
As the field transformation in both cases are strictly analogous,
all results for physical quantities  such as emitted photon flux or
photon statistics can be transposed from one system to the other
with the appropriate choice for $\beta$. In both systems $\beta$
measures the efficiency of the coupling and gives a
measure of the number of emitted photons produced via these
processes. The mechanical oscillation is limited to frequencies in
the GHz regime with maximum amplitudes in the range of less than a
nanometer. At best $\beta_\textrm{mech}$ thus evaluates to
$\beta_\textrm{mech} \leq 10^{-9}$. In contrast, the amplitude of
the apparent cavity length modulation  depends on the frequency and
OPOs are commonly used in the optical frequency range.
Equation (\ref{epsilon approx}) shows that the order of
magnitude of $\epsilon_\textrm{opt}$ will be given by
$\epsilon_\textrm{opt}\simeq \kappa l \sim 10^{-5} l$, therefore the
parameter $\beta_\textrm{opt}$ for the non-linear optical process
can be by many orders of magnitude larger than for the dynamical
Casimir effect based on mechanical motion, e.g. $\beta_\textrm{opt}
\simeq 10 l $ for a pump frequency of $\Omega/2\pi\simeq3.10^{14}$
Hz.

Let us now consider a high finesse cavity ($F\simeq \pi/2\rho >>1$),
with one mirror oscillating at frequency $\Omega$. If the cavity
length is $L_{0}$, a resonant enhancement of the Casimir radiation
will occur when $\Omega=2m \pi c/L_{0}$, i.e. when the oscillation
frequency is chosen to be a multiple integer of the fundamental
cavity mode. The number of photons emitted inside and outside the
cavity via the dynamical Casimir effect after a time $t>>1/\Omega$
has been evaluated within the scattering approach in the
past\cite{Lambrecht1996,Lambrecht1998} and we will just recall the
result here:
\begin{eqnarray}
<N(t)> &\simeq& \frac{2m}{3} \beta_\textrm{opt/mech}^2 (F/\pi)^2  \label{<N(t)> Lambrecht 98} \\
<N^\textrm{out}(t)> &\simeq& \frac{2}{3} \beta_\textrm{opt/mech}^{2}
(F/\pi) \times \frac{\Omega t}{2\pi}. \label{photons}
\end{eqnarray}
These expressions have been obtained far below the oscillation threshold which is given by $\beta_\textrm{opt/mech}=\pi/2F$ and
by assuming a perfect tuning
between $\Omega$ and the cavity modes. The effect of detuning has been studied in detail in \cite{Dodonov1998}.

As we have analogous field transformations for the model OPO system
and the cavity with a single oscillating mirror, we can apply the
above equations to both systems. With the above discussion on the
value of the parameter $\beta$, it is straightforward to discuss the
different orders of magnitude for the dynamical Casimir radiation on
one hand and the Casimir-like photon emission on the other hand. For
a cavity of finesse $F =10^4$ the mechanical motion gives rise to a
photon flux $\frac{<N^{out}(t)>}{t}$ of about $10^{-6}$ photons/s.
For the optical analogue we obtain $\beta_\textrm{opt}\simeq
10^{-6}$, if we consider a laser beam of power $\hbar\Omega\Phi=1$W,
at resonance inside the cavity, pumping a crystal of length $l=0.1
\mu$m over an area $A=10^{-10}$m$^{2}$ at a frequency
$\Omega/2\pi\simeq3.10^{14}$Hz. We then expect an
extracavity radiation of $\frac{<N^\textrm{out}(t)>}{t}\simeq10^{5}$
photons per second excited from vacuum, which largely exceeds
predictions for Casimir radiation in systems based on mechanical
motion. This is simply due to the fact that the model OPO system
avoids direct mechanical motion but uses an apparent modulation of
the cavity length instead. Even though the OPO's
oscillation regime reproducing the dynamical Casimir effect is not
standard, we expect the Casimir-like photons to be emitted by the
usual parametric conversion processes taking place inside the
crystal. The analogue Casimir radiation is thus not an additional
but the only signal emitted by the OPO in this regime. Let us also
underline that our proposal is different from usual OPO experiments
because the non-linear crystal is required to be thinner than the
pumps' wavelength, in order to have a suitable Casimir-like
oscillation regime. This should relax the dispersion and
phase-matching constraints, and permit the simultaneous oscillation
of several parametric resonances.

We finally want to discuss the difference between our predictions,
which imply the existence of a stationary regime for the emitted
photon flux, in comparison with other approaches which predict
exponential growth
\cite{Uhlmann2004,mendonca2005,mendonca2008,Dodonov1995,Schaller2002}.
Obviously, it is of crucial importance for any
experimental observation of the Casimir radiation to know whether
the extracavity photon flux increases linearly or exponentially in
time. To point out the discrepancy and explain its origin we apply
in the reminder of the paper our set-up to the theoretical framework
developed recently in \cite{Uhlmann2004,mendonca2005,mendonca2008}
for cavities with time-dependent refractive media. There the
resulting number of Casimir-like photons inside a perfectly
reflecting cavity takes the form
\begin{equation}
<N(t)>=\sum_{k} \sinh^{2}[r_{k}(t)],
\label{<Nm(t)> cavité pompage
<-}
\end{equation}
where $r_{k}(t)$ represents the squeezing factor for photons emitted
in a given cavity mode $\omega_{k}(t)=k \frac{\pi c}{L(t)}$. Let us
consider the same sinusoidal length change $L(t)$ as before with
$\epsilon << L_{0}$, and use the mean cavity modes $\omega_{k,0}=k
\frac{\pi c}{L_{0}}$. For a modulation frequency $\Omega=2m\pi
c/L_{0}$, pairs of Casimir photons are mainly emitted into the
degenerate cavity mode $\omega_{m,0}=\Omega/2=m\pi c/L_{0}$, with a
squeezing function $r_{m}(t)\simeq \nu_0 t \simeq
\frac{\epsilon}{L_{0}}\frac{\Omega t}{2}$ \cite{mendonca2008}. The
parameter $\nu_0$ measures as $\beta$ the efficiency of photon
creation and corresponds to $\beta = \nu_0 \tau$, where
$\tau=L_{0}/c$ is the time of flight for photons over the mean
cavity length.

In order to account for cavity losses, it has been repeatedly
proposed \cite{mendonca2008,Onofrio06} to introduce a linear damping
rate $\gamma$. If we assume that cavity losses are
mainly due to photons escaping the cavity because of partially
transmitting mirrors, then the linear damping rate should be linked
to the cavity finesse as $\gamma \tau=\frac{\pi }{ F}$. In the
limiting case of a high finesse cavity the
number of emitted photons (\ref{<Nm(t)> cavité
pompage <-}) should be affected by a decay factor $\exp(-\gamma
t)$\cite{mendonca2008}, leading to an exponential growth in the photon emission inside and outside
the cavity in the long time limit
\begin{eqnarray}
<N_{m}(t)> &=& \sinh^{2}(\nu_{0} t) e^{-\gamma t} \simeq \frac{1}{4} e^{(2\nu_{0} - \gamma) t} \\
<N_{m}^\textrm{out}(t)> &=& \sinh^{2}(\nu_{0} t)\big(1-e^{-\gamma t} \big )\simeq \frac{1}{4} e^{2\nu_{0} t}
\label{<Nm(t)>}
\end{eqnarray}

\begin{figure}[tbh]
\begin{center}
\includegraphics[width=9cm]{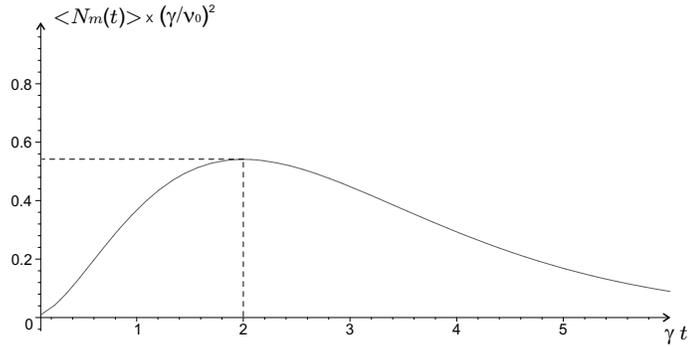}
\caption{Graphic representation of the intracavity photon number
below threshold in the short time limit.} \label{fig2}
\end{center}
\end{figure}

Clearly, equations (\ref{<N(t)> Lambrecht 98}) and (\ref{<Nm(t)>})
 give quantitatively different
predictions for the photon emission rate via the dynamical Casimir
effect. This can however be easily understood by considering the
number of intracavity photons far below threshold, i.e. for
$2\nu_{0}<<\gamma$. In the short time limit $\nu_{0}t<<1$, we can
write $<N_{m}(t)> \simeq (\nu_{0} t)^{2} e^{-\gamma t}$. This
function reaches a maximum value at $t=2/\gamma$, and decreases
afterwards such as shown in figure \ref{fig2}.
The reason for this is that when $\gamma t \rightarrow 2$,
or equivalently when the number of round-trips $t/2\tau$ performed
by photons inside the cavity approaches $F/\pi$, the losses balance
the amplification effect. The oscillating cavity should thus reach
a stationary state with a constant number $<N_{m}>$ of intracavity
Casimir photons given by its maximum value
\begin{eqnarray}
<N_{m}> &\simeq& \frac{1}{e^{2}}
\bigg(\frac{2\nu_{0}}{\gamma}\bigg)^{2}=
\frac{1}{e^{2}}(\beta_\textrm{opt} F/\pi)^{2}. \label{<Nm> below
thr}
\end{eqnarray}
The discrepancy between this result and the one given by equation
(\ref{<N(t)> Lambrecht 98}) now mainly consists in a factor $m$. An
explanation could be that the calculations performed in
\cite{mendonca2008} only account for the contribution of degenerate
pairs of photons emitted in the same mode $\omega_{m,0}=m\frac{\pi
c}{L_{0}}$. In contrast, according to our previous works
\cite{Lambrecht1996,Lambrecht1998}, all pairs of photons
$\big(\omega_{k,0};\omega_{k',0}\big)$ satisfying
$\omega_{k,0}+\omega_{k',0}=\Omega$ ( i.e. $k+k'=2m$ ) should
contribute to the emitted spectrum. When accounting for all
parametric resonances the '$m$' factor should be recovered.

In conclusion we argue that the exponential growth of photon flux is
valid only in the short time limit while our method remains valid in
the long time limit, as it includes right from the beginning a non
unitary reflection coefficient $r$ for the mirrors. Accordingly one
observes a stationary regime with a constant number of Casimir
photons inside the cavity. The difference between this method and
the introduction of an energy loss $\gamma$ is that only by using
reflection coefficients the phase relations for the fields are
explicitly taken into account and automatically respected.

While the dynamical Casimir effect is still unobserved today, an
intermediate step in order to achieve this task  would be to use the
above OPO model system which obeys rigorously the same field
transformations as a cavity with a single oscillating mirror. This
could give considerable additional insight into the dynamical
Casimir effect especially as far as experimental questions are
concerned. It already permits to test a number of important
predictions for the dynamical Casimir effect in easily achievable
conditions. In particular it would allow to test whether
the present prediction for the emitted photon flux is correct, i.e.
that the system evolves into a stationary state with a constant
number of Casimir photons inside the cavity and an extracavity
radiation growing linearly in time or if these quantities grow
exponentially. The clarification of this point would be of greatest
importance for any experimental set-up aiming at observing the
dynamical Casimir radiation due to direct mechanical motion.

\acknowledgments The authors thank Serge Reynaud for valuable
discussions and the ESF Research Networking Programme CASIMIR
(www.casimir-network.com) for providing excellent opportunities for
discussions on the Casimir effect and related topics.

\end{document}